# Convex Optimal Power Flow Based on Power Injection-based Equations and Its Application in Bipolar DC Distribution Network

Yiyao Zhou, Qianggang Wang, *Member, IEEE*, Yuan Chi, *Member, IEEE*, Jianquan Liao, *Member, IEEE*, Tao Huang, *Member, IEEE*, Niancheng Zhou, *Member, IEEE*, Xiaolong Xu and Xuefei Zhang

*Abstract*—Optimal power flow (OPF) is a fundamental tool for analyzing the characteristics of bipolar DC distribution network (DCDN). However, existing OPF models face challenges in reflecting the power distribution and exchange of bipolar DCDN directly since its decision variables are voltage and current. This paper addresses this issue by establishing a convex OPF model that can be used for the planning and operation of bipolar DCDN. First, the power flow characteristics of bipolar DCDN are revealed through power injection-based equations, upon which the original OPF model is established. Next, the original OPF model undergoes a transformation into a convex OPF model based on second-order cone programming (SOCP) through variable substitution, second-order cone relaxation, McCormick relaxation, and first-order Taylor expansion, respectively. Finally, the sequence bound tightening algorithm (STBA) is employed to tighten the boundaries of McCormick envelopes in each iteration to ensure the exactness of the convex OPF model. The effectiveness of this novel OPF model for bipolar DCDN is verified through two case studies, i.e., capacity configuration of distributed generation (DG) and operation optimization of bipolar DCDN.

NOMENCLATURE

**Indices:**

| | |
|---|---|
| $\varphi$ | Pole index of +, o and - representing positive, neutral, and negative pole |
| $\rho$ | Port index of p, n and b representing +~o, o~- and +~- connection |

**Sets:**

| | |
|---|---|
| $N$ | Set of all nodes |
| $N^+$ | Set of nodes other than the slack node |
| $\vartheta$ | Set of branches |
| $N_{VB}/N_{DG}$ | Set of VBs and DGs |

**Parameters:**

| | |
|---|---|
| $r_{ij}$ | The resistance of the branch $i$~$j$ |
| $U_{ps}/U_{ns}$ | Output voltage of p and n for VB |
| $z_{\rho,j}, i_{\rho,j}$ and $p_{\rho,j}$ | Coefficients of load model |
| $\alpha, \beta$ and $\gamma$ | Weight coefficients of cost |
| $a_{VB}, b_{VB}$ and $c_{VB}$ | Cost coefficients of VB |
| $u_G/c_G$ | Unit cost of DG for investment and compensation |
| $c_{loss}$ | Cost of unit loss |
| $P_{L,\rho}^{base}/U_\rho^{base}$ | Nominal load power and voltage of $\rho$ |
| $\bar{U}_{\varphi,i}/\underline{U}_{\varphi,i}$ | The upper and lower limits of $U_{\varphi,i}$ |
| $\bar{I}_{\varphi,ij}/\underline{I}_{\varphi,ij}$ | The upper and lower limits of $I_{\varphi,ij}$ |
| $\bar{P}_{G,\rho,i}/\underline{P}_{G,\rho,i}$ | The upper and lower limits of $P_{G,\rho,j}$ |
| $\delta$ | The upper limit of the voltage unbalance |
| $\bar{V}_{\varphi,i}/\underline{V}_{\varphi,i}$ | The upper and lower limits of $V_{\varphi,i}$ |
| $\bar{L}_{\varphi,ij}/\underline{L}_{\varphi,ij}$ | The upper and lower limits of $L_{\varphi,ij}$ |

**Variables:**

| | |
|---|---|
| $P_{\varphi,ij}/I_{\varphi,ij}$ | Power and current of $i$~$j$ for $\varphi$ |
| $U_{\varphi,i}$ | Voltage of $i$ for $\varphi$ |
| $P_{L,\varphi,j}$ | The net load power of $j$ for $\varphi$ |
| $P_{L,\rho,j}$ | The load power of $j$ for $\rho$ |
| $P_{G,\rho,j}$ | DG output of $j$ for $\rho$ |
| $V_{\varphi,i}/L_{\varphi,ij}$ | The square of $U_{\varphi,i}$ and $I_{\varphi,ij}$ |
| $P_{VB,\rho,0}$ | Output power of VB for $\rho$ |
| $P_{VB,\rho,j,t}$ | Output power of VB for $\rho$ at $t$ |

## I. INTRODUCTION

The rapid advancement of power electronics technology and the increasing prevalence of DC source loads have created a new opportunity for the growth of DC distribution network (DCDN) [1]. The structure of DCDN can be classified into two groups, i.e., unipolar and bipolar. The bipolar DCDN has one more neutral conductor than unipolar DCDN. Therefore, bipolar DCDN offers greater reliability, transmission capacity, and grid flexibility compared to unipolar structures [2], but its complex network characteristics of bipolar DCDN pose challenges for the planning and operation [3]. Despite these difficulties, the greater development potential of bipolar

DCDNs makes them a promising area of research.

The optimal power flow (OPF) is a vital tool for the operation analysis of the DCDN, which can reflect the most economical power flow distribution state of the system under physical and operational constraints [4]. The OPF model for DCDN is inherently non-convex due to the presence of nonlinear terms in both power flow and operational constraints [5]. There are three common approaches to address this issue: 1) adopting heuristic algorithm to solve the non-convex OPF model [6]. 2) adopting iterative algorithm to seek the local optimal solution of the non-convex OPF model [7] and 3) using commercial software to solve the convex OPF model relaxed by convexification technique [8].

The first approach, which uses heuristic algorithms such as genetic algorithm and particle swarm optimization, does not provide global optimal solution of the OPF model as these algorithms trade optimality and accuracy for convenient implementation (such as no requirement for any transformation of the original non-convex model [9]) and speed. For the second method, iterative algorithms, such as the interior point method and Newton-Raphson method, are usually adopted to seek the optimal solution by following the boundaries of constraints [10]. These methods do not provide global optimality [5], although empirical evidence suggests that the Newton-Raphson method is quite successful in practice [7].

Compared with the former algorithm-wise methods, the convexification method has drawn more research interest in recent years since a convex OPF model promises a globally optimal solution under certain conditions [10]. Existing research is dedicated to transforming the original OPF model into a convex model by relaxation and approximation techniques that rigorously guarantee the following two properties during the convex transformation: 1) the exactness of the convex OPF model and 2) the reducibility of the solution for the convex OPF model. A variety of techniques have been utilized in research to transform the original OPF model into a convex model, including second-order cone relaxation [8], quadratic convex approximation [11], Taylor first-order expansion [12], etc. Among them, second-order cone relaxation can not only guarantee the existence and uniqueness of the optimal solution, but also achieve exact relaxation with any topological structure and operation mode [13]. Therefore, the OPF model of DCDN formulated as second-order cone programming (SOCP) is preferred in research. In addition, the original OPF model can also be transformed into other formers, such as semidefinite programming model (SDP) [14], mixed-integer quadratic programming model (MIQP) [15], mixed-integer linear programming model (MILP) [16], etc. The successful application of the aforementioned models in the unipolar DCDN provides a solid foundation for their potential application in the bipolar DCDN. Under ideal conditions where the positive and negative poles are entirely symmetrical, the bipolar DCDN can be decoupled into unipolar networks. However, such ideal conditions rarely exist in an electrical engineering practice due to the presence of non-negligible asymmetric disturbances caused by the sources and loads connected to the positive and negative poles. Consequently, the OPF problem of bipolar DCDN is more intricate than that of the unipolar DCDN. The approaches of OPF for unipolar DCDN is difficult to directly be implemented since the complex characteristics of coupled network and asymmetrical operation for bipolar DCDN.

The current injection-based equations can precisely describe the structure and reveal characteristics of bipolar DCDN [17]. The OPF model based on current injection equations can be decomposed into a series of subproblems, which are expressed as quadratic programming (QP) problems that utilize the Jacobian matrix. As the iteration of the Jacobian matrix continues, the subproblems are updated and gradually approach the optimal solution of the original OPF model [17]. For the bipolar DCDN with constant power loads (CPLs), the original OPF problem can be decomposed into convex and nonconvex QP problems, which can be solved by nonlinear programming solvers [18]. But this method is computationally expensive for a high accuracy of the optimal solution. To address the problem of solution efficiency caused by CPLs, researchers shift their focus toward the field of convex optimization [19-21]. After the current equations of CPLs are relaxed and approximated, the original OPF model based on current injection equations becomes a convex OPF model, which is widely used to optimize network loss [17], generation cost [18], locational marginal price [22], voltage unbalance [23] and asymmetrical loads [24].

TABLE I
SUMMARY OF RESEARCH ON OPF IN DCDN

| Power Flow Model | Literature | Decision Variable Forms | Modeling Method | Applicable Scenarios |
|---|---|---|---|---|
| Current injection-based equations | [18] | Voltage and current | QP | Symmetrical/Asymmetrical bipolar DCDN |
| | [19] | | MIQP | |
| | [21] | | Recursive conic approximation | |
| | [22] | | Linear approximation | |
| Power injection-based equations | [8], [13] | Square of voltage and power | SOCP | Symmetrical bipolar DCDN Unipolar DCDN |
| | [14] | | SDP | |
| | [16] | | MILP | |
| | [15] | Voltage and power | MIQP | |

The current injection-based equations are linear, but the bilinear terms are introduced into the objective function, i.e., the product of voltage and current, which is nonconvex and generates a weak duality between the objective function and constraints [25]. Existing current injection-based OPF models cannot directly reflect the power exchange between the bipolar

DCDN and the external grid, particularly the AC grid. We can refer to the solution of the unipolar DCDN, i.e. adopting power injection-based equations to formulate OPF model of the bipolar DCDN. Table I summarizes the relevant research on OPF of DCDN from the aspects of power flow model, decision variables, modeling method, and application scenarios. It can be seen that there is no research on OPF of bipolar DCDN with the power as a decision variable and reflecting power flow with power injection-based equations.

The power injection-based equations are both nonlinear and non-convex. The convexification method can be used to solve the problem [18]. This paper proposes a convex OPF model of bipolar DCDN based on power injection equations. The convex OPF model guarantees exact relaxation and efficient computation, which are the central problems in pursuing the convexification method [8]. Furthermore, this model is applicable to both planning and operation stages of bipolar DCDN. The main contributions of this paper are as follows:

1) A novel original OPF model of bipolar DCDN is proposed, which formulates the power flow model with power injection-based equations. Additionally, power is selected as a decision variable, which illustrates the proposed model is a promising OPF model for power optimization in bipolar DCDN.

2) A method for exact relaxation and approximation of the original OPF model for bipolar DCDN is proposed. The proposed method transforms the original OPF model into a convex OPF model in SOCP form, ensuring the uniqueness and reducibility of the optimal solution.

3) The sequence bound tightening algorithm (SBTA) was employed to ensure the tightness of the McCormick envelopes. This further minimizes the error that results from the convex conversion of the original OPF model.

The rest of the paper is organized as follows: the basic OPF model of bipolar DCDN is formulated in Section II. In Section III, convex relaxation and linear approximation are further applied to the basic OPF model. Then, the exactness of convex OPF model is analyzed and the algorithm for reducing the error is proposed in Section IV. In Section V the results of our numerical study are shown and analyzed. Section VI concludes the paper with the final remarks.

## II. PROBLEM FORMULATION

In this section, the original OPF model of the bipolar DCDN is established based on the power injection-based equations to avoid the bilinear objective function and the nonlinear current equations of CPLs.

### A. Power flow model

The bipolar DCDN has positive, neutral and negative conductors, which are denoted by index of +, $o$ and – respectively. The voltage balancers (VBs), sources and loads are connected to three conductors through the impedance network. The specific structure is shown in Fig. 1. Node 0 denotes the slack node with a fixed voltage, which is connected to VB. Let $i\sim j$ denotes a pair $\{i, j\}$ with electrical connection in $N^+$, and $i\sim j \in \vartheta$.

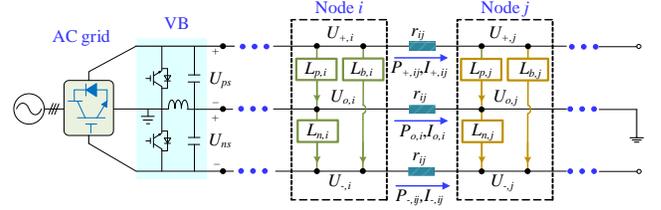

Fig. 1. The structure of bipolar DCDN.

The power flow equations of bipolar DCDN are governed the following physical laws:

1) Branch $i\sim j$ is governed by Ohm's law.

$$I_{\varphi,ij} = \frac{U_{\varphi,i} - U_{\varphi,j}}{r_{ij}}, i \sim j \in \vartheta \quad (1)$$

2) Power of branch $i\sim j$ is calculated by node voltage and branch current.

$$P_{\varphi,ij} = U_{\varphi,i} I_{\varphi,ij}, i \sim j \in \vartheta \quad (2)$$

3) Power injection-based equations of node $j$.

$$P_{L,\varphi,j} = \sum_{i:i\sim j}(P_{\varphi,ij} - I_{\varphi,ij}^2 r_{ij}) - \sum_{j:j\sim k} P_{\varphi,jk}, j \in N^+ \quad (3)$$

where subscript $k$ denotes the downstream of $j$.

$$\begin{cases} P_{L,+,j} = (P_{L,p,j} + P_{L,b,j}) - (P_{G,p,j} + P_{G,b,j}) \\ P_{L,o,j} = (P_{L,n,j} - P_{L,p,j}) - (P_{G,n,j} - P_{G,p,j}), j \in N^+ \\ P_{L,-,j} = (P_{G,p,j} + P_{G,b,j}) - (P_{L,n,j} + P_{L,b,j}) \end{cases} \quad (4)$$

The static load model for DC grids can be expressed using the constant resistance, constant current, and constant power models, similar to the AC load model [26].

$$P_{L,\rho,j} = P_{L,\rho}^{base}[z_{\rho,j}(\frac{U_{\rho,j}}{U_\rho^{base}})^2 + i_{\rho,j}\frac{U_{\rho,j}}{U_\rho^{base}} + p_{\rho,j}], j \in N^+ \quad (5)$$

where $z_{\rho,j}$, $i_{\rho,j}$ and $p_{\rho,j}$ must sum to one.

The transformation of port voltage and node voltage can be expressed as (6).

$$\begin{bmatrix} U_{p,j} \\ U_{n,j} \\ U_{b,j} \end{bmatrix} = \begin{bmatrix} 1 & -1 & 0 \\ 0 & 1 & -1 \\ 1 & 0 & -1 \end{bmatrix} \begin{bmatrix} U_{+,j} \\ U_{o,j} \\ U_{-,j} \end{bmatrix}, j \in N^+ \quad (6)$$

### B. Optimal power flow problem

The objective of the OPF problem can be expressed as a function of decision variables, as shown in (7), which is usually the operation cost and loss of network [27].

$$F = f(\boldsymbol{x}_{\text{ori}}) \quad (7)$$

where vector $\boldsymbol{x}_{\text{ori}}$ are the decision variables of the original OPF model, and $\boldsymbol{x}_{\text{ori}} = \{U_{\varphi,i}, I_{\varphi,ij}, P_{\varphi,ij}\}$.

The constraints of the original OPF model include physical constraints, security constraints, and DG capacity constraints. The physical constraints have been described in Section II-A, and the remaining constraints are as follows:

1) Security constraints

$$\begin{cases} U_{\varphi,0} = U_{\varphi,0}^{ref} \\ \underline{U}_{\varphi,i} \leq U_{\varphi,i} \leq \overline{U}_{\varphi,i} \quad , i \in N^+ \end{cases} \quad (8)$$

$$\underline{I}_{\varphi,ij} \leq I_{\varphi,ij} \leq \overline{I}_{\varphi,ij} \quad , i \sim j \in \vartheta \quad (9)$$

$$VUF_i = \frac{|U_{p,i} - U_{n,i}|}{0.5(U_{p,i} + U_{n,i})} \leq \delta \quad , i \in N^+ \quad (10)$$

The voltage security constraint is (8), which specifies that the voltage of the slack node is fixed as $U_{\varphi,0}^{ref}$, and the voltage of $j$ for $\varphi$ cannot exceed its upper and lower limits. The current security constraint is (9), which specifies that the current of $i \sim j$ for $\varphi$ cannot exceed its upper and lower limits. The voltage unbalance security constraint is (10), which specifies that the voltage unbalance degree of $i$ cannot exceed its upper limit.

2) DG capacity constraints

$$\underline{P}_{G,\rho,i} \leq P_{G,\rho,i} \leq \overline{P}_{G,\rho,i} \quad , i \in N^+ \quad (11)$$

Equation (11) specifies that the DG output of $i$ for $\rho$ cannot exceed its upper and lower limits.

This paper focuses solely on the basic components of bipolar DCDN and does not include power electronic equipment such as DC transformers, power flow controllers, DC power springs, etc.

To summarize, the original OPF model can be formulated as (12), and abbreviate original OPF model by ori-OPF.

$$\text{ori-OPF:} \quad \begin{aligned} & \min F \\ & \text{over } x_B \\ & \text{subject to:} \\ & (1)-(6),(8)-(11) \end{aligned} \quad (12)$$

The decision variable of the ori-OPF cannot be the port voltage, as the transformation matrix between port voltage and node voltage in (6) is singular [18].

The ori-OPF is based on the following assumptions:
1) The bipolar DCDN ($N$, $\vartheta$) is connected.
2) The reference direction of the current is the direction of the arrow in Fig. 1.
3) The lines of bipolar DCDN ($N$, $\vartheta$) are lossy [8].

## III. CONVEX RELAXATION AND LINEAR APPROXIMATION

The ori-OPF is a nonlinear and non-convex optimization problem. This section illustrates the application of convex relaxation and linear approximation on the ori-OPF with a detailed process depicted in Fig. 2.

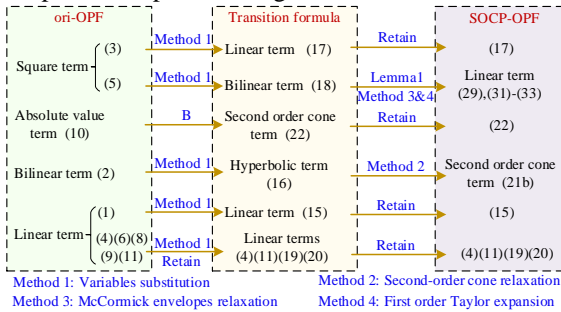

Fig. 2. The processing of convex relaxation and linear approximation.

### A. Relaxation of the square term

The variables substitution is adopted to relax the square terms in (3) and (5).

The following slack variables are introduced

$$\begin{cases} L_{\varphi,ij} = I_{\varphi,ij}^2 \quad , i \sim j \in \vartheta \\ V_{\varphi,i} = U_{\varphi,i}^2 \quad , i \in N^+ \end{cases} \quad (13)$$

For each branch $i \sim j$, the matrix A is rank one and positive semidefinite, the matrix A is expressed as

$$A = \begin{bmatrix} U_{\varphi,j} \\ I_{\varphi,ij} \end{bmatrix} \begin{bmatrix} U_{\varphi,j} & I_{\varphi,ij} \end{bmatrix} = \begin{bmatrix} V_{\varphi,i} & P_{\varphi,ij} \\ P_{\varphi,ij} & L_{\varphi,ij} \end{bmatrix} \quad (14)$$

Then, (1)-(3), (5), and (8)-(9) are transformed into the following equations.

$$V_{\varphi,j} = V_{\varphi,i} - 2r_{ij}P_{\varphi,ij} + r_{ij}^2 L_{\varphi,ij} \quad , i \sim j \in \vartheta \quad (15)$$

$$P_{\varphi,ij}^2 = L_{\varphi,ij}V_{\varphi,j} \quad , i \sim j \in \vartheta \quad (16)$$

$$P_{L,\varphi,j} = \sum_{i: i \sim j}(P_{\varphi,ij} - L_{\varphi,ij}r_{ij}) - \sum_{j: j \sim k} P_{\varphi,jk} \quad , j \in N^+ \quad (17)$$

$$P_{L,p,j} = P_{L,p}^{base} \left[ \frac{z_{p,j}(V_{+,j} + V_{o,j} - 2U_{+,j}U_{o,j})}{(U_p^{base})^2} \right.$$
$$\left. + i_{p,j}\frac{U_{+,j} - U_{o,j}}{U_p^{base}} + p_{p,j} \right] \quad , j \in N^+ \quad (18a)$$

$$P_{L,n,j} = P_{L,n}^{base} \left[ \frac{z_{n,j}(V_{o,j} + V_{-,j} - 2U_{o,j}U_{-,j})}{(U_n^{base})^2} \right.$$
$$\left. + i_{n,j}\frac{U_{o,j} - U_{-,j}}{U_n^{base}} + p_{n,j} \right] \quad , j \in N^+ \quad (18b)$$

$$P_{L,b,j} = P_{L,b}^{base} \left[ \frac{z_{b,j}(V_{+,j} + V_{-,j} - 2U_{+,j}U_{-,j})}{(U_b^{base})^2} \right.$$
$$\left. + i_{b,j}\frac{U_{+,j} - U_{-,j}}{U_\rho^{base}} + p_{b,j} \right] \quad , j \in N^+ \quad (18c)$$

$$\begin{cases} V_{\varphi,0} = V_{\varphi,0}^{ref} \\ \underline{V}_{\varphi,i} \leq V_{\varphi,i} \leq \overline{V}_{\varphi,i} \quad , i \in N^+ \end{cases} \quad (19)$$

$$\underline{L}_{\varphi,ij} \leq L_{\varphi,ij} \leq \overline{L}_{\varphi,ij} \quad , i \sim j \in \vartheta \quad (20)$$

In (15)-(20), (16) and (17) are obtained by substituting (13) into (2) and (3), (18) is obtained by substituting (6) and (13) into (5), (19) and (20) are obtained by substituting (13) into (8) and (9). (15) is derived in Appendix A.

A new square term appears in (16) after variable substitution. Considering the positive semidefinite property of the matrix A, (16) can be relaxed into (21a), and further written as the second-order cone form, i.e., (21b).

$$P_{\varphi,ij}^2 \leq L_{\varphi,ij}V_{\varphi,i} \quad , i \sim j \in \vartheta \quad (21a)$$

$$\left\| \begin{matrix} 2P_{\varphi,ij} \\ L_{\varphi,ij} - V_{\varphi,i} \end{matrix} \right\|_2 \leq L_{\varphi,ij} + V_{\varphi,i} \quad , i \sim j \in \vartheta \quad (21b)$$

## B. Relaxation of the absolute value term

The second-order cone relaxation is adopted to deal with the absolute value term in (10), and the relaxed voltage imbalance security constraint is (22), which is derived in Appendix B.

$$\begin{cases} \dfrac{V_{+,i}^2 + V_{-,i}^2}{a^2 - 2} \leq V_{+,i} V_{-,i} \Leftrightarrow \left\| \begin{array}{c} 2V_{+,i}/\sqrt{a^2-2} \\ 2V_{-,i}/\sqrt{a^2-2} \\ V_{+,i} - V_{-,i} \end{array} \right\|_2 \leq V_{+,i} + V_{-,i} \\ a = \dfrac{2 + 0.5\delta^2}{1 - 0.25\delta^2} \end{cases} \quad (22)$$

## C. Relaxation of the bilinear term

There are still bilinear items in (18) after variables substitution. Despite $U_{o,j}$ and $V_{o,j}$ can be ignored since the neutral conductor is grounded, the bilinear term $U_{+,j}U_{-,j}$ in (18c) still needs to be relaxed. Before that, we propose *Lemma* 1.

*Lemma* 1: Given branch $i\sim j$ and node $i$, if the matrix A is rank one and satisfies (13), and the voltage of slack node satisfies (23).

$$\begin{cases} U_{+,0} = \sqrt{V_{+,0}} \\ U_{-,0} = -\sqrt{V_{-,0}} \end{cases} \quad (23)$$

Then, $U_{\varphi,i}$ and $I_{\varphi,ij}$ can be reduced by (24) and (25), respectively.

$$\begin{cases} U_{+,i} = \sqrt{V_{+,i}} \ , i \in N^+ \\ I_{+,ij} = \sqrt{L_{+,ij}} \ , i \sim j \in \vartheta \end{cases} \quad (24)$$

$$\begin{cases} U_{-,i} = \sqrt{V_{-,i}} \ , i \in N^+ \\ I_{-,ij} = \sqrt{L_{-,ij}} \ , i \sim j \in \vartheta \end{cases} \quad (25)$$

The *Lemma* 1 is proved in Appendix C.

*Lemma* 1 guarantees the reducibility of the optimal solution after substituting the decision variable of ori-OPF.

Ignoring $U_{o,j}$ and $V_{o,j}$ in (18), and use Lemma 1 to transform (18) into (26).

$$P_{L,p,j} = P_{L,p,j}^{base} [\dfrac{z_{p,j} V_{+,j}}{(U_p^{base})^2} + i_{p,j} \dfrac{\sqrt{V_{+,j}}}{U_p^{base}} + p_{p,j}] \ , j \in N^+ \quad (26a)$$

$$P_{L,n,j} = P_{L,n,j}^{base} [\dfrac{z_{n,j} V_{-,j}}{(U_n^{base})^2} - i_{n,j} \dfrac{\sqrt{V_{-,j}}}{U_n^{base}} + p_{n,j}] \ , j \in N^+ \quad (26b)$$

$$P_{L,b,j} = P_{L,b,j}^{base} [\dfrac{z_{b,j}(V_{+,j} + V_{-,j} + 2\sqrt{V_{+,j} V_{-,j}})}{(U_b^{base})^2} \\ + i_{b,j} \dfrac{\sqrt{V_{+,j}} + \sqrt{V_{-,j}}}{U_\rho^{base}} + p_{b,j}] \ , j \in N^+ \quad (26c)$$

McCormick relaxation is adopted to deal with the bilinear term in (26c). For arbitrary bilinear term $xy$, $x \in [\underline{x}, \overline{x}]$, $y \in [\underline{y}, \overline{y}]$, can be mathematically described as (27) [28].

$$\text{McCormick}(x, y, [\underline{x}, \overline{x}], [\underline{y}, \overline{y}]) =$$

$$\text{Upper Bounds} \begin{cases} w \geq \underline{x} y + x\underline{y} - \underline{x}\underline{y} \\ w \geq \overline{x} y + x\overline{y} - \overline{x}\overline{y} \end{cases} \quad (27)$$

$$\text{Lower Bounds} \begin{cases} w \leq \underline{x} y + x\overline{y} - \underline{x}\overline{y} \\ w \leq \overline{x} y + x\underline{y} - \overline{x}\underline{y} \end{cases}$$

Introduce $v_j = V_{+,j} V_{-,j}$, $V_{-,j} \in [\underline{V}_{-,j}, \overline{V}_{-,j}]$, (26c) can be rewritten as (28), and (29) is introduced to constrain $v_j$.

$$P_{L,b,j} = P_{L,b,j}^{base} [\dfrac{z_{b,j}(V_{+,j} + V_{-,j} + 2\sqrt{v_j})}{(U_b^{base})^2} \\ + i_{b,j} \dfrac{\sqrt{V_{+,j}} + \sqrt{V_{-,j}}}{U_\rho^{base}} + p_{b,j}] \ , j \in N^+ \quad (28)$$

$$\text{McCormick}(V_{+,j}, V_{-,j}, [\underline{V}_{+,j}, \overline{V}_{+,j}], [\underline{V}_{-,j}, \overline{V}_{-,j}]) \ , j \in N^+ \quad (29)$$

## D. Linear approximation based on Taylor expansion

The first-order Taylor expansion is adopted to approximate the root terms in (26a), (26b), and (28), which are introduced by the relaxation of the bilinear terms. The first-order Taylor expansion can be mathematically described as (30).

$$\sqrt{z} = \dfrac{z}{2\sqrt{z_0}} + \dfrac{\sqrt{z_0}}{2} \quad (30)$$

Since the per unit value of $U_{+,j}$ and $U_{-,j}$ is around 1.0 (actually within the interval [0.95,1.05]), then the per unit value of $V_{+,j}$, $V_{-,j}$ and $v_j$ is also around 1.0. Therefore, the Taylor expansion can be performed on $\sqrt{V_{+,j}}$, $\sqrt{V_{-,j}}$ and $\sqrt{v_j}$ at 1.0 and retain the first-order term, i.e., (31)-(33).

$$P_{L,p,j} = P_{L,p,j}^{base} [\dfrac{z_{p,j} V_{+,j}}{(U_p^{base})^2} + i_{p,j} \dfrac{V_{+,j}+1}{2U_p^{base}} + p_{p,j}] \ , j \in N^+ \quad (31)$$

$$P_{L,n,j} = P_{L,n,j}^{base} [\dfrac{z_{n,j} V_{-,j}}{(U_n^{base})^2} - i_{n,j} \dfrac{V_{-,j}+1}{2U_n^{base}} + p_{n,j}] \ , j \in N^+ \quad (32)$$

$$P_{L,n,j} = P_{L,n,j}^{base} [\dfrac{z_{n,j} V_{-,j}}{(U_n^{base})^2} - i_{n,j} \dfrac{V_{-,j}+1}{2U_n^{base}} + p_{n,j}] \ , j \in N^+ \quad (33)$$

## E. OPF model based on second-order cone programming

After convex relaxation and linear approximation, the ori-OPF can be reformulated into the SOCP form in (34), which can be abbreviated by SOCP-OPF.

$$\begin{aligned} \text{SOCP-OPF:} \quad & \min f(x_{\text{SOCP}}) \\ & \text{over } x_{\text{SOCP}} = \{V_{\varphi,i}, L_{\varphi,ij}, P_{\varphi,ij}, v_i\} \\ & \text{subject to:} \\ & (4), (11), (15), (17), (19), (20), (21b) \\ & , (22), (29), (31) - (33) \end{aligned} \quad (34)$$

SOCP-OPF can be efficiently solved by commercial software and it possesses a unique optimal solution, as guaranteed by *Theorem* 1.

*Theorem* 1: If SOCP-OPF is both convex and exact, then it can have at most one optimal solution.

The *Theorem* 1 is proved in Appendix D.

*Theorem* 1 demonstrates that a convex and exact SOCP is a sufficient condition for the existence and uniqueness of the optimal solution. The constraints of SOCP-OPF are all second-order cone constraints or linear constraints, which guarantee the convex property of SOCP-OPF. The exactness of relaxation and approximation will be analyzed and improved in Section IV.

## IV. EXACTNESS ANALYSIS AND ERROR REDUCTION

Four techniques are adopted to transform the ori-OPF into SOCP-OPF. However, relaxation errors may be generated by the techniques other than variable substitution. This section analyzes and reduces such relaxation errors.

*Errors caused by linearization*: linearizing (26a), (26b), and (28) may introduce errors.

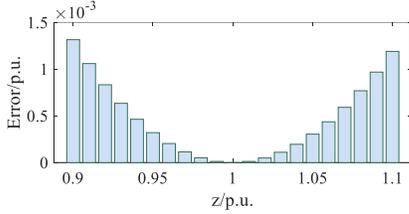

Fig. 3. The error curve for approximation.

Define the approximation error as $\mu(z)$. Fig. 3. shows that the curve of $\mu(z)$ on the order of $10^{-3}$, which is within the acceptable range.

$$\mu(z) = \left| \frac{1}{2}z + \frac{1}{2} - \sqrt{z} \right| \qquad (35)$$

*Errors caused by relaxation*: relaxing (10) and (16) may introduce errors.

Relaxations of equality/inequality on (16) and (10) to inequalities in (21b) and (22) may lead to errors in the model [29], since (21b) and (22) only have lower bounds but have no upper bounds. To ensure the tightness of these relaxations, we can establish upper bounds on (21b) and (22) and bring the lower and the upper bounds closer. This is achieved by substituting bilinear terms $L_{\varphi,ij}V_{\varphi,i}$ and $V_{+,i}V_{-,i}$ with $W_{\varphi,ij}$ and $v_i$ in (21b) and (22) and introducing McCormick envelops relaxation. The second-order cone constraints (21b) and (22) will be redefined as (37) and (38), and McCormick envelops constraints (39) and (29) are used to ensure that (37) and (38) have upper and lower bounds.

$$\left\| \begin{matrix} 2P_{\varphi,ij} \\ W_{\varphi,ij} - 1 \end{matrix} \right\|_2 \leq W_{\varphi,ij} + 1 \;,\; i \sim j \in \vartheta \qquad (37)$$

$$\left\| \begin{matrix} 2V_{+,i}/\sqrt{a^2-2} \\ 2V_{-,i}/\sqrt{a^2-2} \\ v_i - 1 \end{matrix} \right\|_2 \leq v_i + 1 \;,\; i \in N^+ \qquad (38)$$

McCormick$(L_{\varphi,ij}, V_{\varphi,i}, [\underline{L}_{\varphi,ij}, \overline{L}_{\varphi,ij}], [\underline{V}_{\varphi,i}, \overline{V}_{\varphi,i}])$ , $i \sim j \in \vartheta$ (39)

Then, SOCP-OPF is further expressed in the form of (40), which can be abbreviated by McSOCP-OPF.

McSOCP-OPF': $\min f(x_{\text{McSOCP}})$

over $x_{\text{McSOCP}} = \{V_{\varphi,i}, L_{\varphi,ij}, W_{\varphi,ij}, P_{\varphi,ij}, v_i\}$

subject to: (40)

(4), (11), (15), (17), (19), (20),

, (29), (31) – (33), (37) – (39)

Compared with SOCP-OPF, McSOCP-OPF supplements the upper bounds to the second-order cone constraints. But the McCormick envelopes in McSOCP-OPF may not be tight. Sequence bound tightening algorithm (STBA) is adopted to ensure the tightness of the McCormick envelopes [29]-[31]. The pseudocode of STBA is presented in Algorithm 1.

| **Algorithm 1**: Sequence bound tightening algorithm |
|---|
| 1: **Inputs**: tolerance $\varepsilon$, sequence $k$, step length $d$ |
| 2: **Initialize**: $k \leftarrow 1$; $\underline{L}_{\varphi,ij}^{ini} \leftarrow \underline{L}_{\varphi,ij}$, $\overline{L}_{\varphi,ij}^{ini} \leftarrow \overline{L}_{\varphi,ij}$ $\underline{V}_{\varphi,i}^{ini} \leftarrow \underline{V}_{\varphi,i}$, $\overline{V}_{\varphi,i}^{ini} \leftarrow \overline{V}_{\varphi,i}$ |
| 3: **repeat** |
| 4:   **if** $k=1$ **then** |
| 5:     Solve (40), and get $V_{\varphi,i}^k, L_{\varphi,ij}^k, W_{\varphi,ij}^k, v_i^k$ |
| 6:     Calculate the relaxation error $\lambda_k = \max\left\{ \left\| \frac{W_{\varphi,ij}^k - L_{\varphi,ij}^k V_{\varphi,i}^k}{W_{\varphi,ij}^k} \right\|, \left\| \frac{v_i^k - V_{+,i}^k V_{-,i}^k}{v_i^k} \right\| \right\}$ |
| 7:   **else** |
| 8:     $\underline{L}_{\varphi,ij} \leftarrow \max\{(1-r^k)L_{\varphi,ij}^k, \underline{L}_{\varphi,ij}^{ini}\}$ |
| 9:     $\overline{L}_{\varphi,ij} \leftarrow \min\{(1+r^k)L_{\varphi,ij}^k, \overline{L}_{\varphi,ij}^{ini}\}$ |
| 10:     $\underline{V}_{\varphi,ij} \leftarrow \max\{(1-r^k)V_{\varphi,i}^k, \underline{V}_{\varphi,ij}^{ini}\}$ |
| 11:     $\overline{V}_{\varphi,ij} \leftarrow \min\{(1+r^k)V_{\varphi,i}^k, \overline{V}_{\varphi,ij}^{ini}\}$ |
| 12:   **end if** |
| 13:   $k=k+1$ |
| 14: **until** $\lambda_k \leq \varepsilon$ |

In the first iteration, the McSOCP-OPF is solved according to the initial bounds of $L_{\varphi,ij}$ and $V_{\varphi,i}$. In subsequent iterations, the bounds of $L_{\varphi,ij}$ and $V_{\varphi,i}$ are updated with the optimal solution obtained in the $k^{th}$ iteration. The feasible area of $L_{\varphi,ij}$ and $V_{\varphi,i}$ will be tightened in each iteration. Thereby, the tightness of McCormick envelopes constraints is improved. The iterative process is terminated until the error index $\lambda_k$ is no more than the given tolerance $\varepsilon$.

## V. CASE STUDIES

The application effect of McSOCP-OPF in the planning and operation of bipolar DCDN is analyzed in this section. Two application scenarios of McSOCP-OPF are set up, i.e., 1) DG capacity configuration and 2) operation optimization of bipolar DCDC with DG.

The simulations are conducted in MATLAB 2018a on a 64-bit computer with a 3.00 GHz CPU and 16.0 GB RAM. The optimization models are programmed on the YALMIP platform, and the McSOCP-OPF is solved by the GUROBI 9.5 solver.

## A. Optimization configuration of DG capacity

In case A, the effectiveness of McSOCP-OPF and STBA is verified by configuring DG capacity on a 5-node feeder. It is noteworthy that case A does not involve the problem of DG location.

Fig. 4. illustrates a 5-node feeder connected to the AC grid through VSC. A VB is installed in node 0, which can be equivalent to two sources with a capacity of 1kW connected to $p$ and $n$ ports [18]. Constant power loads with a total capacity of 18.8kw are connected to nodes 1 to 4 with three connection methods, i.e., $p$, $n$, and $b$, and baseline loads are shown in Table II. Line resistance, nominal voltage, voltage deviation limit and voltage unbalance limit are 0.1Ω, ±400V, ±5% and 3%, respectively. The positive, neutral, and negative currents do not exceed [-2,2] A, [-1,1] A, [-2,2] A, respectively. $d$=0.02, and $\varepsilon$=10$^{-6}$.

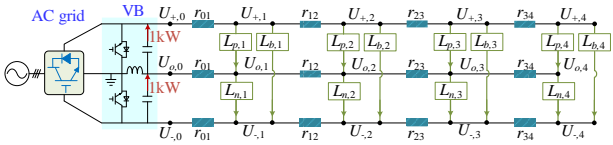

Fig. 4. Topology of the 5-node feeder.

TABLE II
PARAMETERS OF LOAD IN 5-NODE FEEDER

| Connection | Load 1 (W) | Load 2 (W) | Load 3 (W) | Load 4 (W) |
|---|---|---|---|---|
| $p$ | 500 | 1000 | 800 | 900 |
| $n$ | 1000 | 600 | 700 | 400 |
| $b$ | 1200 | 1500 | 1200 | 1000 |

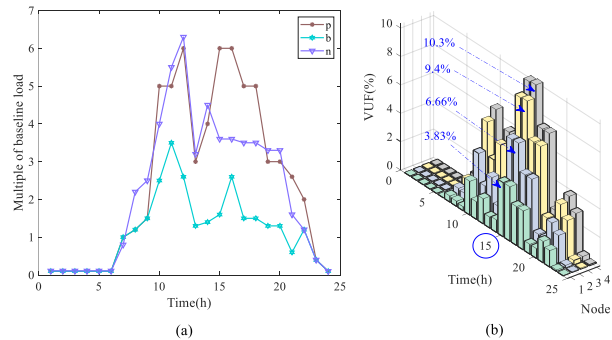

Fig. 5. Load variation and voltage unbalance degree distribution of 5-node feeder. (a) Load variation. (b) Voltage unbalance degree distribution.

Fig. 5(a) presents the multiple relationship between the real load and the baseline load for each period. The power flow calculation method in [17] is adopted to calculate the profile of voltage unbalance for the 5-node feeder, and the results are shown in Fig. 5(b). At 15h, the 5-node feeder experiences the most severe unbalanced voltage problem. The load for $p$ is 6 times the baseline load, while the loads for $n$ and $b$ are 2.6 times and 3.5 times the baseline load, respectively. As a result, the voltage unbalance for nodes 1-4 are 3.83%, 6.66%, 9.4%, and 10.3%, respectively.

We need to configure enough capacity DG on this feeder to optimize the unbalanced voltage, despite the extreme load condition of 16h. The decision variables of this problem are $x_{sce1} = \{x_{SOCP}, |P_{G,\rho,j}, P_{VB,\rho,0}\}$. The objective is expressed as (41).

Fig. 6. shows the results of DG configuration and the corresponding voltage unbalance profile of the 5-node feeder at 15h. Despite being with a heavy load, the voltage unbalance of this feeder is improved effectively. The voltage unbalance on this feeder does not exceed 0.0564%, which is significantly lower than the upper limit of 3%.

$$f(x_{sce1}) = \alpha c_G \sum_{\rho=p}^{b} \sum_{j \in N^+} P_{G,\rho,j} + \beta c_{loss} \sum_{\varphi=+}^{-} \sum_{i \sim j \in \vartheta} L_{\varphi,ij} r_{ij} \\ + \gamma \sum_{\rho=p}^{b} (a_{VB} P_{VB,\rho,0}^2 + b_{VB} P_{VB,\rho,0} + c_{CB}) \quad (41)$$

where $\alpha=\beta=\gamma=1/3$, $c_G$=603.19 \$/kW, $c_{loss}$=0.5\$/kWh, $a_{VB}$=8×10$^{-5}$ \$/kW$^2$h $b_{VB}$=0.08 \$/kWh and $c_{VB}$=0.

Table III presents a comparison of the operation of the 5-node feeder before and after configuring DG. It clearly shows that the demand for electricity from the main network significantly decreases after DG configuration, which can be attributed to the improved operational efficiency that results in a significant reduction of network losses.

TABLE III
EFFECT OF OPERATION FOR THE 5-NODE FEEDER

| | Network loss (W) | Power of VB(W) | |
|---|---|---|---|
| | | $p$ | $n$ |
| Before configuration | 2717.6 | 1000 | 1000 |
| After configuration | 2.6122 | 552.5 | 552.6 |

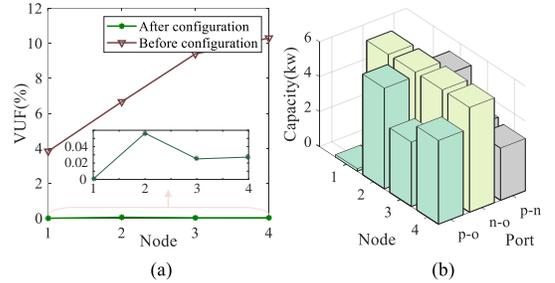

Fig. 6. The result and effect of DG configuration for 5-node feeder. (a) Voltage unbalance degree distribution at 16h. (b) DG capacity of 5-node feeder.

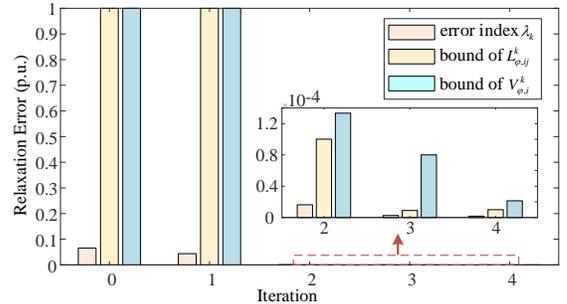

Fig. 7. Error on $\lambda_k$ calculation and $L_{\varphi,ij}^k / V_{\varphi,i}^k$ bound decrements over the iterations in SBTA.

Fig. 7 presents the performance of STBA. STBA plays a key role in tightening the McCormick envelopes by significantly reduce the range of $L_{\varphi,ij}^k$ and $V_{\varphi,i}^k$ in each iteration. The relaxation error has been successfully minimized to below the tolerance, i.e., 10$^{-6}$, with only 4 iterations. The cumulative calculation time is about 4.5665s. Additionally, the errors of second-order cone relaxation in (37) and (38) are also reduced since the McCormick envelope is tightened. It is noteworthy that the calculation time of STBA is inversely related to the $r$. However, a larger $r$ increases the likelihood of McSOCP-OPF missing the optimal solution or entering an oscillating iterative process.

## B. Optimal operation of bipolar DCDN with DG

The application effect of McSOCP-OPF in the operation optimization of bipolar DCDN with DG is verified in a modified IEEE 33-node test system [18] illustrated in Fig. 8 as Case B.

The line resistance of the test system is the same as the original test system. The nominal voltage and rated capacity of the test system are ±3kV and 1MVA. The load profile is shown in Fig. 9., and the load connection is the same as the [18]. DGs are 500kW connected to $p$ and $n$ at node 22. VBs are 500kW at nodes 1, 14, and 31. The positive, neutral, and negative currents do not exceed [-10,10] A, [-1,1] A, [-10,10] A, respectively. The other parameters are the same as in Case A except for $\varepsilon$, whose value is $10^{-3}$.

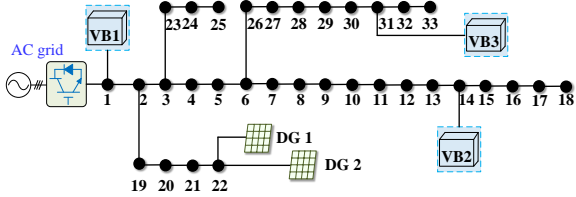

Fig. 8. The modified IEEE 33-node test system.

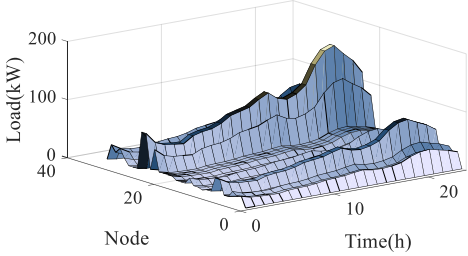

Fig. 9. Load profile of test system.

The purpose of Case A is to optimize voltage profile to reduce voltage unbalance, which is achieved by promoting power-voltage balance. The decision variables of this problem are $x_{sce1} = \{x_{SOCP}, |P_{G,\rho,j,t}, P_{VB,\rho,j,t}\}$. The objective function is expressed as (42).

$$f(x_{sce2}) = \alpha c_G \sum_{t=1}^{24} \sum_{\rho=p}^{b} \sum_{j \in N_G} P_{G,\rho,j,t} + \beta c_{loss} \sum_{t=1}^{24} \sum_{\varphi=+i\sim}^{-} \sum_{j \in \vartheta} L_{\varphi,ij,t} r_{ij}$$
$$+ \gamma \sum_{t=1}^{24} \sum_{\rho=p}^{b} \sum_{j \in N_{VB}} (a_{VB} P_{VB,\rho,j,t}^2 + b_{VB} P_{VB,\rho,j,t} + c_{CB})$$
(42)

where $\alpha=0.8$, $\beta=\gamma=0.1$, $c_G=0.8$ \$/kW. The other parameters are the same as (41).

Fig. 10. and 11 are the optimization results of output power for DGs and VBs, respectively. It can be seen that: 1) Due to the positive loads being 20% more than the negative loads at each node, the output power of DGs and VBs connected to $p$ is obviously more than that of $n$. 2) The loads of the test system are heavy during 19-23h (the average load of each node reaches 450.16kW during this period), which causes more output power of DGs and VBs than during other periods. 3) Nodes 24 and 25 are close to VB1, and the loads of $p$ is heavy than other ports, which leads to the output power of VB1 being the highest among all VBs connected to $p$.

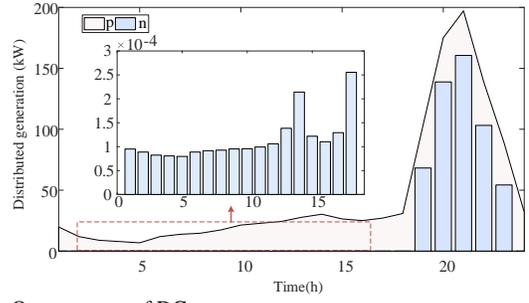

Fig. 10. Output power of DGs.

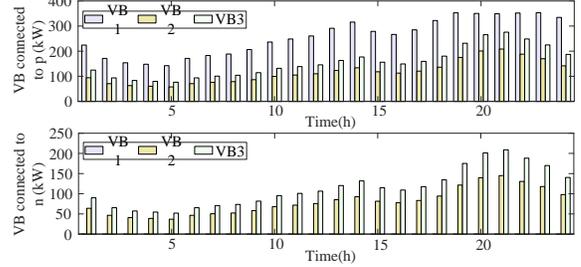

Fig. 11. Output power of VBs.

Fig. 12. illustrates the optimization results of voltage, from which the following conclusions can be drawn: 1) The maximum positive, negative, and neutral voltage deviations are 0.37%, 0.23%, and 0.59%, respectively, and the maximum voltage unbalance is only 1.78%. This indicates that the optimization of the output power of DGs and VBs has a positive effect on reducing the voltage deviation and unbalance of the test system. 2) The heavier loads lead to greater voltage unbalance and voltage differences between positive and negative poles. This phenomenon is particularly obvious during the 19-23h period and underscores the importance of asymmetric load as a key cause of voltage unbalance. 3) nodes 24 and 25 present the same phenomenon as descried in (2), mainly due to the fact that heavier loads exacerbate the asymmetry between positive and negative.

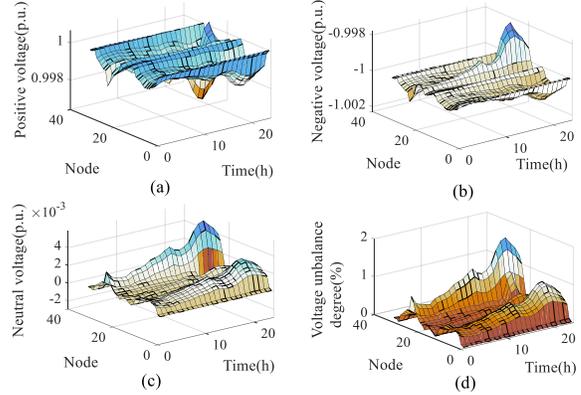

Fig. 12. Optimization results of bipolar DCDN. (a) Positive voltage profile. (b) Negative voltage profile. (c) Neutral voltage profile. (d) Voltage unbalance degree profile.

Fig. 13. shows the errors of both the second-order cone relaxation and McCormick relaxation in the McSOCP-OPF. All errors are within the required accuracy threshold ($\varepsilon$ is $10^{-3}$ in this case). To further enhance the exactness of SOCP-OPF, it is possible to decrease $\varepsilon$ and $r$. Nonetheless, this could negatively impact the solution efficiency.

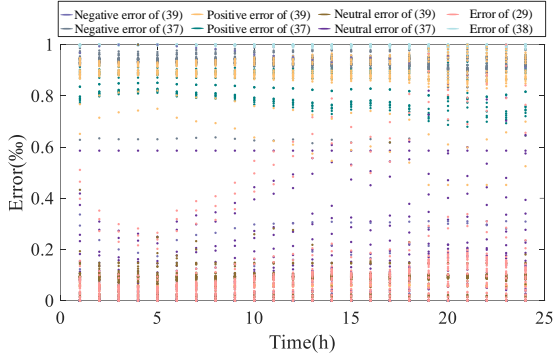

Fig. 13. The errors analysis of second-order cone relaxation and McCormick relaxation.

## VI. Conclusion

This paper proposes a convex OPF model considering power injection-based equations in the bipolar DCDN. The specific conclusions are as follows:

1) As a novel OPF model with the power as a decision variable, its convex property enables efficient solutions using commercial software. The calculation results can directly present the flow and exchange of power in the bipolar DCDN.

2) The exactness of the proposed model is achieved by STAB, which can successfully reduce errors in each iteration. Specifically, the relaxation error can be reduced by nearly 1000 times in a single iteration.

3) The proposed model can effectively optimize DG capacity, network loss, DG output, voltage deviation, and voltage unbalance, thereby demonstrating its potential for wide application in the planning and operation of the bipolar DCDN.

## Appendix

### A. Derivation of (15)

Rewrite equation (1) as

$$U_{\varphi,j} = U_{\varphi,i} - I_{\varphi,ij} r_{ij} \ , j \in N^+$$

Square both sides of the equation, get

$$U_{\varphi,j}^2 = U_{\varphi,i}^2 - 2U_{\varphi,i} I_{\varphi,ij} r_{ij} + I_{\varphi,ij}^2 r_{ij}^2 \ , j \in N^+$$

(15) is obtained by substituting (2) and (13) into (2) above equation. This completes the derivation of (15).

### B. Derivation of (22)

Substitute (6) into (10), get

$$\frac{|U_{+,i} + U_{-,i} - 2U_{o,i}|}{0.5(U_{+,i} - U_{-,i})} \leq \delta \ , i \in N^+$$

Obviously, both sides of "≤" are non-negative, and $U_{+,i}-U_{-,i} \neq 0$. Therefore, both sides of the inequality can be squared and rewritten as

$$(U_{+,i} + U_{-,i} - 2U_{o,i})^2 \leq 0.25\delta^2 (U_{+,i} - U_{-,i})^2 \ , i \in N^+$$

Substitute (13) into above inequation, get

$$(V_{+,i} + V_{-,i})(1 - 0.25\delta^2) + (2 + 0.5\delta^2)V_{+,i}V_{-,i}$$
$$\leq 4U_{o,i}(U_{+,i} + U_{-,i} - U_{o,i}) \ , i \in N^+$$

Since $0 < \delta < 1$, the above inequality can be transformed into

$$\begin{cases} V_{+,i} + V_{-,i} \leq \dfrac{f}{(1-0.25\delta^2)} - aU_{+,i}U_{-,i} \\ f = 4U_{o,i}(U_{+,i} + U_{-,i} - U_{o,i}) \end{cases} , i \in N^+$$

Since the neutral conductor of bipolar DCDN is grounded, $U_{o,j}$ is both close to zero. Therefore, $f \approx 0$ and $f \ll -aU_{+,i}U_{-,i}$. Under this assumption, the above inequality can be transformed into

$$V_{+,i} + V_{-,i} \leq -aU_{+,i}U_{-,i} \ , i \in N^+$$

Square both sides of the inequality, get

$$V_{+,i}^2 + V_{-,i}^2 \leq (a^2 - 2)V_{+,i}V_{-,i} \ , i \in N^+$$

Since $a = \dfrac{2 + 0.5\delta^2}{1 - 0.25\delta^2} = 2 + \dfrac{\delta^2}{1 - 0.25\delta^2} > 2$, above inequality can be written as the second-order cone form in (22). This completes the derivation of (22).

### C. Proof of Lemma 1

If (13) is satisfied in arbitrary branch $i \sim j$ and node $i$, and the matrix A is rank one. Then

$$P_{\varphi,ij}^2 = V_{\varphi,i} L_{\varphi,ij}$$

Since $P_{\varphi,ij} > 0$, then

$$P_{\varphi,ij} = \sqrt{V_{\varphi,i}} \sqrt{L_{\varphi,ij}} \ \vee \ P_{\varphi,ij} = (-\sqrt{V_{\varphi,i}})(-\sqrt{L_{\varphi,ij}})$$

Assume the positive/negative voltage and current are opposite to (24) and (25), i.e.

$$\begin{cases} U_{+,i} = -\sqrt{V_{+,i}} \ , i \in N^+ \\ I_{+,ij} = -\sqrt{L_{+,ij}} \ , i \sim j \in \vartheta \end{cases}, \begin{cases} U_{-,i} = \sqrt{V_{-,i}} \ , i \in N^+ \\ I_{-,ij} = \sqrt{L_{-,ij}} \ , i \sim j \in \vartheta \end{cases}$$

Then the positive current flows from $j$ to $i$ ($j>i$), i.e., $U_{+,j} > U_{+,i}$. Since $(N, \vartheta)$ is connected, such propagation can continue and eventually one has $U_{+,0} < 0$. This contradicts with (23).

Similarly, the contradiction between negative voltage/current and (23) can also be deduced.

Therefore, for arbitrary branch $i \sim j$ and node $i$, $U_{\varphi,j}$ and $I_{\varphi,ij}$ can only be reduced by (24) and (25). This completes the proof of the *Lemma* 1.

### D. Proof of Theorem 1

Assume that SOCP-OPF is convex, exact, and has at least one solution. Let $\tilde{x} = (\tilde{V}, \tilde{L}, \tilde{P}, \tilde{v})$ and $\hat{x} = (\hat{V}, \hat{L}, \hat{P}, \hat{v})$ be two optimal solution of SOCP-OPF. If $\tilde{x} = \hat{x}$, then *Theorem* 1 holds.

Let $x = (\tilde{x} + \hat{x})/2$ is the average value of $\tilde{x}$ and $\hat{x}$. Then, $x$ is also the solution of SOCP-OPF. Therefore, $x$, $\tilde{x}$ and $\hat{x}$ all satisfy rank[A]=1, i.e.,

$$\begin{cases} \tilde{V}_{\varphi,i} \tilde{L}_{\varphi,ij} = \tilde{P}_{\varphi,ij}^2 \\ \hat{V}_{\varphi,i} \hat{L}_{\varphi,ij} = \hat{P}_{\varphi,ij}^2 \ , i \sim j \in \vartheta \\ V_{\varphi,i} L_{\varphi,ij} = P_{\varphi,ij}^2 \end{cases}$$

Substitute $V_{\varphi,i} = (\tilde{V}_{\varphi,i} + \hat{V}_{\varphi,i})/2$ and $L_{\varphi,ij} = (\tilde{L}_{\varphi,ij} + \hat{L}_{\varphi,ij})/2$ into above equation, get

$$\tilde{V}_{\varphi,i} \hat{L}_{\varphi,ij} + \hat{V}_{\varphi,i} \tilde{L}_{\varphi,ij} = 2\tilde{P}_{\varphi,ij} \hat{P}_{\varphi,ij} \ , i \sim j \in \vartheta$$

Since both $\tilde{P}_{\varphi,ij}$ and $\hat{P}_{\varphi,ij}$ are positive, the above equation is equivalent to

$$\tilde{V}_{\varphi,i}\hat{L}_{\varphi,ij} + \hat{V}_{\varphi,i}\tilde{L}_{\varphi,ij} = 2\sqrt{\tilde{V}_{\varphi,i}\tilde{L}_{\varphi,ij}\hat{V}_{\varphi,i}\hat{L}_{\varphi,ij}} \Leftrightarrow$$

$$(\sqrt{\tilde{V}_{\varphi,i}\hat{L}_{\varphi,ij}} - \sqrt{\hat{V}_{\varphi,i}\tilde{L}_{\varphi,ij}})^2 = 0 \quad, i \sim j \in \vartheta$$

Therefore

$$\tilde{V}_{\varphi,i}\hat{L}_{\varphi,ij} = \hat{V}_{\varphi,i}\tilde{L}_{\varphi,ij} \Leftrightarrow \frac{\tilde{V}_{\varphi,i}}{\hat{V}_{\varphi,i}} = \frac{\tilde{L}_{\varphi,ij}}{\hat{L}_{\varphi,ij}} \quad, i \sim j \in \vartheta$$

In addition, $x$, $\tilde{x}$ and $\hat{x}$ all satisfy (15), combined with the above equation, get:

$$\frac{\tilde{V}_{\varphi,i}}{\hat{V}_{\varphi,i}} = \frac{\tilde{V}_{\varphi,j} + 2r_{ij} - r_{ij}^2\tilde{L}_{\varphi,ij}}{\hat{V}_{\varphi,j} + 2r_{ij} - r_{ij}^2\hat{L}_{\varphi,ij}} = \frac{\tilde{L}_{\varphi,ij}}{\hat{L}_{\varphi,ij}} \quad, i \sim j \in \vartheta$$

Let $k = \hat{V}_{\varphi,i}/\tilde{V}_{\varphi,i} = \tilde{L}_{\varphi,ij}/\hat{L}_{\varphi,ij}$, $i \sim j \in \vartheta$, then

$$\tilde{V}_{\varphi,j} + 2r_{ij} - r_{ij}^2\tilde{L}_{\varphi,ij} = k(\hat{V}_{\varphi,j} + 2r_{ij} - r_{ij}^2\hat{L}_{\varphi,ij}) \Rightarrow$$

$$\tilde{V}_{\varphi,j} - k\hat{V}_{\varphi,j} + 2r_{ij}(1-k) - r_{ij}^2(\tilde{L}_{\varphi,ij} - k\hat{L}_{\varphi,ij}) = 0$$

Since $\tilde{L}_{\varphi,ij} - k\hat{L}_{\varphi,ij} = 0$ and $\tilde{V}_{\varphi,j} - k\hat{V}_{\varphi,j} = 0$, then

$$2r_{ij}(1-k) = 0 \Rightarrow k = 1$$

Therefore, $\hat{V}_{\varphi,i} = \tilde{V}_{\varphi,i}$, $\tilde{L}_{\varphi,ij} = \hat{L}_{\varphi,ij}$, $i \sim j \in \vartheta$. Then, $\tilde{P}_{\varphi,ij}^2 = \hat{P}_{\varphi,ij}^2 \Rightarrow \tilde{P}_{\varphi,ij} = \hat{P}_{\varphi,ij}$. $\hat{v}_i = \tilde{v}_i$ obviously holds since $\hat{v}_i$ and $\tilde{v}_i$ can be expressed by $\hat{V}_{\varphi,i}$ and $\tilde{V}_{\varphi,i}$, respectively. This completes the proof of the *Theorem* 1.